\documentclass[12pt]{iopart}
\usepackage{graphicx}
\usepackage{amssymb} 

\begin{document}

\title[ATLAS Heavy Ion Physics Prospects]{Heavy Ion Physics Prospects with the ATLAS Detector at the LHC}
\author{N.~Grau, \textit{for the ATLAS Collaboration}}
\address{Columbia University, Nevis Laboratories\\
         Irvington, NY, USA, 10533}
\ead{ncgrau@nevis.columbia.edu}

\begin{abstract}
The next great energy frontier in Relativistic Heavy Ion Collisions is quickly approaching
with the completion of the Large Hadron Collider and the ATLAS experiment is poised to make
important contributions in understanding QCD matter at extreme conditions. While
designed for high-$p_T$ measurements in high-energy $p+p$ collisions, the detector is well
suited to study many aspects of heavy ion collisions from bulk phenomena to high-$p_T$ and
heavy flavor physics. With its large and finely segmented electromagnetic and
hadronic calorimeters, the ATLAS detector excels in measurements of photons and jets,
observables of great interest at the LHC. In this talk, we highlight the performance of the
ATLAS detector for Pb+Pb collisions at the LHC with special emphasis on a key feature of 
the ATLAS physics program: jet and direct photon measurements.
\end{abstract}
\submitto{\JPG}

\section{Introduction}\label{sec:intro}
The advent of an era where two relativistic heavy ion physics colliders are running simultaneously
at the LHC and at RHIC is upon us. The RHIC program has moved from its initial 
discovery phase into making detailed measurements to further understand the 
strongly-interacting quark-gluon plasma formed in $Au+Au$ collisions. The LHC heavy ion program 
provides an opportunity to perform complementary measurements at higher collision energy 
and to study the plasma with a different temperature and lifetime. Together RHIC and the LHC will 
be a focused, two-pronged attack on understanding QCD matter in extreme conditions.

The RHIC program has brought new insights on aspects of the entire collision evolution 
from the initial state parton distribution functions of the nuclei, through thermalization 
of the fireball, and the subsequent hadronization of the medium. Questions about the role of saturation 
physics came to light with the first measurements of charged particle multiplicity\cite{Back:2000gw}. 
The large value of the measured elliptic flow, $v_2$, for all particles up to and including 
heavy flavor mesons together with hydrodynamical calculations indicate that the matter 
produced is strongly interacting with a low shear viscosity-to-entropy density ratio, a perfect 
fluid, instead of a dilute gas of partons\cite{WhitePapers}. $J/\psi$ suppression was 
observed at RHIC. But the quantitatively similar suppression of SPS and RHIC data suggests that 
additional physics, such as recombination, which enhance the $J/\psi$ signal at RHIC is 
required\cite{Adare:2006ns}. Jet 
quenching was discovered as a suppression of high-$p_T$ hadrons\cite{Adcox:2001jp} and confirmed by the 
away-side suppression from two-particle azimuthal correlations\cite{Adler:2002tq}. Novel structures,
such as the ``shoulder''\cite{Adler:2005ee} and the ``ridge''\cite{Putschke:2006zg} have been measured 
in heavy ion 
collisions. Even with this large amount of data on jet energy loss and its effect on the medium, the
exact nature of energy loss is poorly understood. Single particle measurements ($R_{AA}$) are apparently 
insensitive to details of energy loss\cite{Renk:2006pw} as are two-particle correlations, being 
predominatly due to punch-through and tangential emission\cite{RenkPuncthrough,TangentialJets}.

\begin{figure}[bt]
\begin{center}
\includegraphics[width=0.65\linewidth]{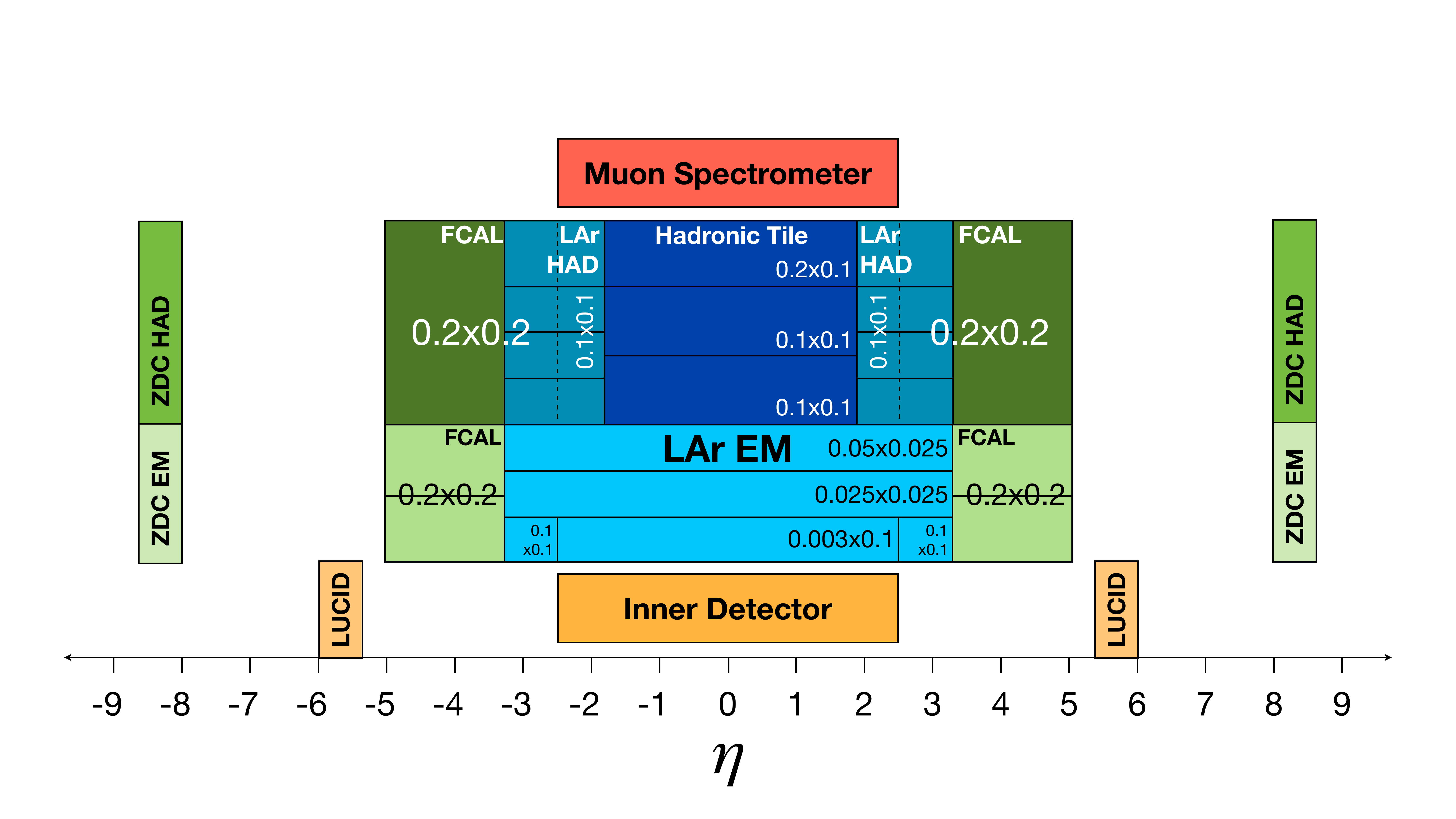}
\end{center}
\caption{The $\eta$ view of the different subdetectors of the ATLAS, all subdetectors
cover the full 2$\pi$ in azimuth. Tracking and muon detection extends to $|\eta|<$2.5. Both the
electromagnetic and hadronic calorimeters cover $|\eta|<$ 5 and are longitudinally segmented
with the typical $\Delta\eta$x$\Delta\phi$ segmentation indicated.}\label{fig:ATLASAcceptance}
\end{figure}

Measurements at the LHC should elucidate these
issues that are currently not well understood at RHIC. ``Day-1'' measurements such as the charged 
particle multiplicity will provide key constraints to saturation physics.
Measurements of the collective flow of particles are crucial 
to determine if a perfect fluid state exists at the LHC. Quarkonia 
measurements, both bottomonium and charmonium, will give more detailed insight on the effects 
contributing to quarkonium suppression. Full jet reconstruction will reduce biases from 
energy loss, since the energy should be radiated predominantly close to the jet 
direction\cite{Salgado:2003rv} and be reconstructed with the jet energy. 
The ATLAS prospects of global physics 
measurements\cite{PeterQM08} and quarkonia\cite{SashaQM08} capabilites are detailed elsewhere 
in these proceedings.  The focus of this contribution is on measurements of jets 
and photons, those that will utilize the strenghts of the ATLAS detector, in order
to understand the details of energy loss of hard scattered partons in the medium.

\section{Heavy Ion Physics Prospects}
The ATLAS detector\cite{ATLASTDR} is poised to make important contributions to the LHC heavy ion
program in the coming $Pb+Pb$ runs. Although designed for high-$p_{T}$ measurements of $p+p$ collisions 
at 14 TeV, the detector has been shown to be well suited to perform heavy ion measurements even
at the extreme edge of currently predicted particle multiplicities ($dN/d\eta \sim 3000$)\cite{LoI}.
The ATLAS detector consists of inner tracking chambers followed by electromagnetic then hadronic 
calorimetery and finally a muon spectrometer. The ATLAS detector covers the full $2\pi$ in azimuth 
and Fig.~\ref{fig:ATLASAcceptance} shows the $\eta$ acceptance.

The unique feature of the ATLAS detector is its calorimetry. It is both electromagnetic and hadronic
and covers 10 units of $\eta$, unprecedented coverage for relativistic heavy ion experiments.
The notable feature of the calorimeter is its longitudinal segmentation with varying 
$\Delta\eta$x$\Delta\phi$ segment sizes indicated in Fig.~\ref{fig:ATLASAcceptance}.
Of particular 
importance is the first longitudinal electromagnetic segment. It is composed of strips in 
$\eta$ with a typical width of 0.003 units in $\eta$ and extends to $|\eta| \lesssim $ 2.5 units. It was 
designed to measure $H\rightarrow\gamma\gamma$ events and to reject di-jet events. 
The importance of this layer for photon measurements is discussed in Section~\ref{sec:photons}.

\subsection{Jet Physics in ATLAS}\label{sec:jets}

Jet performance in heavy ion events has been studied for two complementary jet reconstruction
algorithms, the seeded cone algorithm and the Fast-$k_T$ algorithm\cite{Cacciari:2005hq}. These algorithms 
perform jet reconstruction on 0.1x0.1 $\Delta\eta$x$\Delta\phi$ towers built from energy sums of the 
electromagnetic and hadronic calorimeters. For the seeded cone algorithm, the segment- and
$\eta$-dependent $\langle E_T\rangle$ is subtracted prior to jet reconstruction. 
For the Fast-$k_T$ algorithm jet reconstruction is performed directly on the full energy 
calorimeter towers. A set of discriminant variables is used to distinguish 
real jets from background jets composed of underlying event energy. After discrimination, the 
background jets are used to determine the background energy to be subtracted from the real
jets.

The performace for jet reconstruction is evaluated by embedding entire PYTHIA di-jet
events into unquenched HIJING events.
All reconstructed jets were
compared to the truth jets defined as jets reconstructed from the final state, 
generated particles.

\begin{figure}[tb]
\begin{center}
\includegraphics[width=0.45\linewidth]{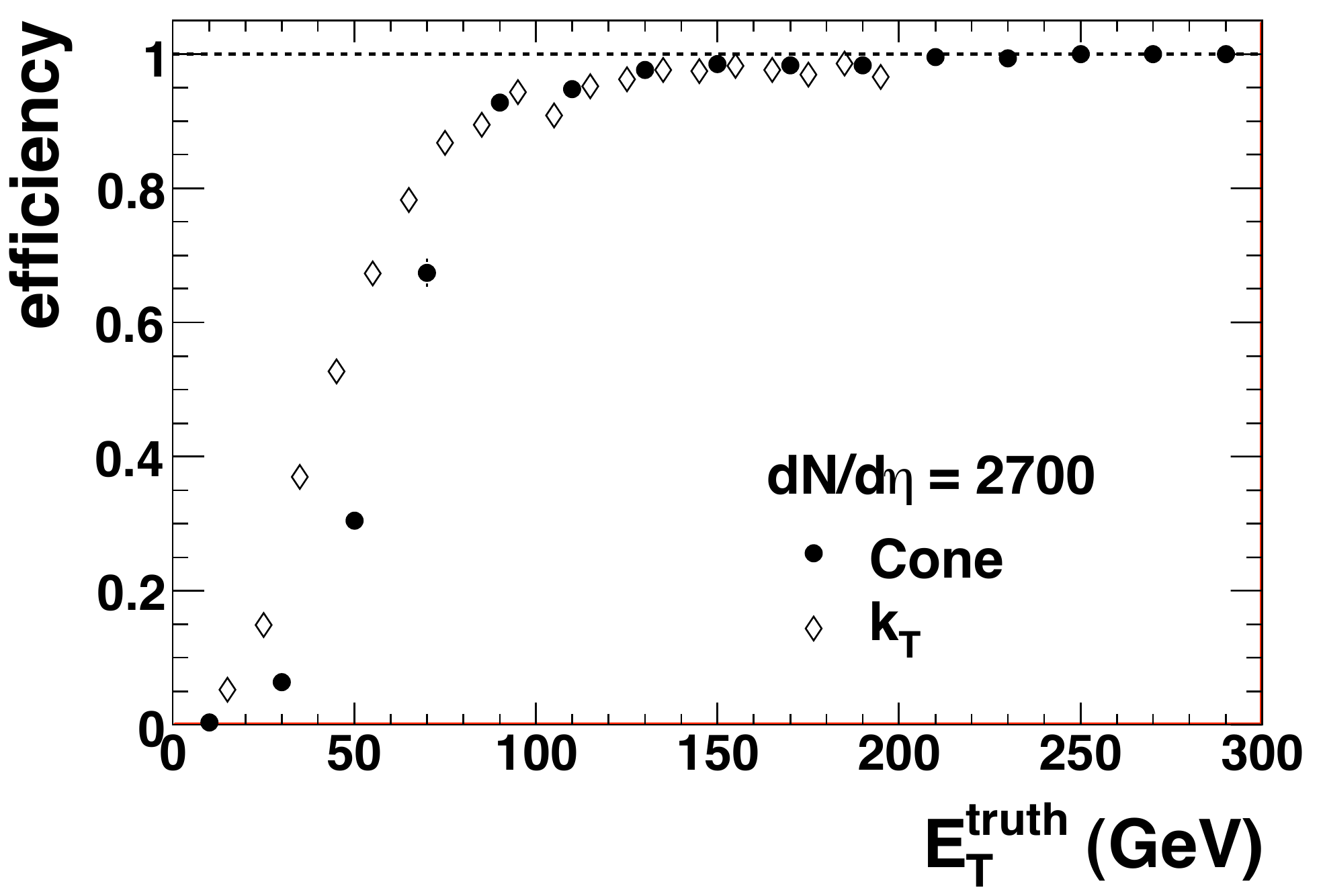}
\includegraphics[width=0.45\linewidth]{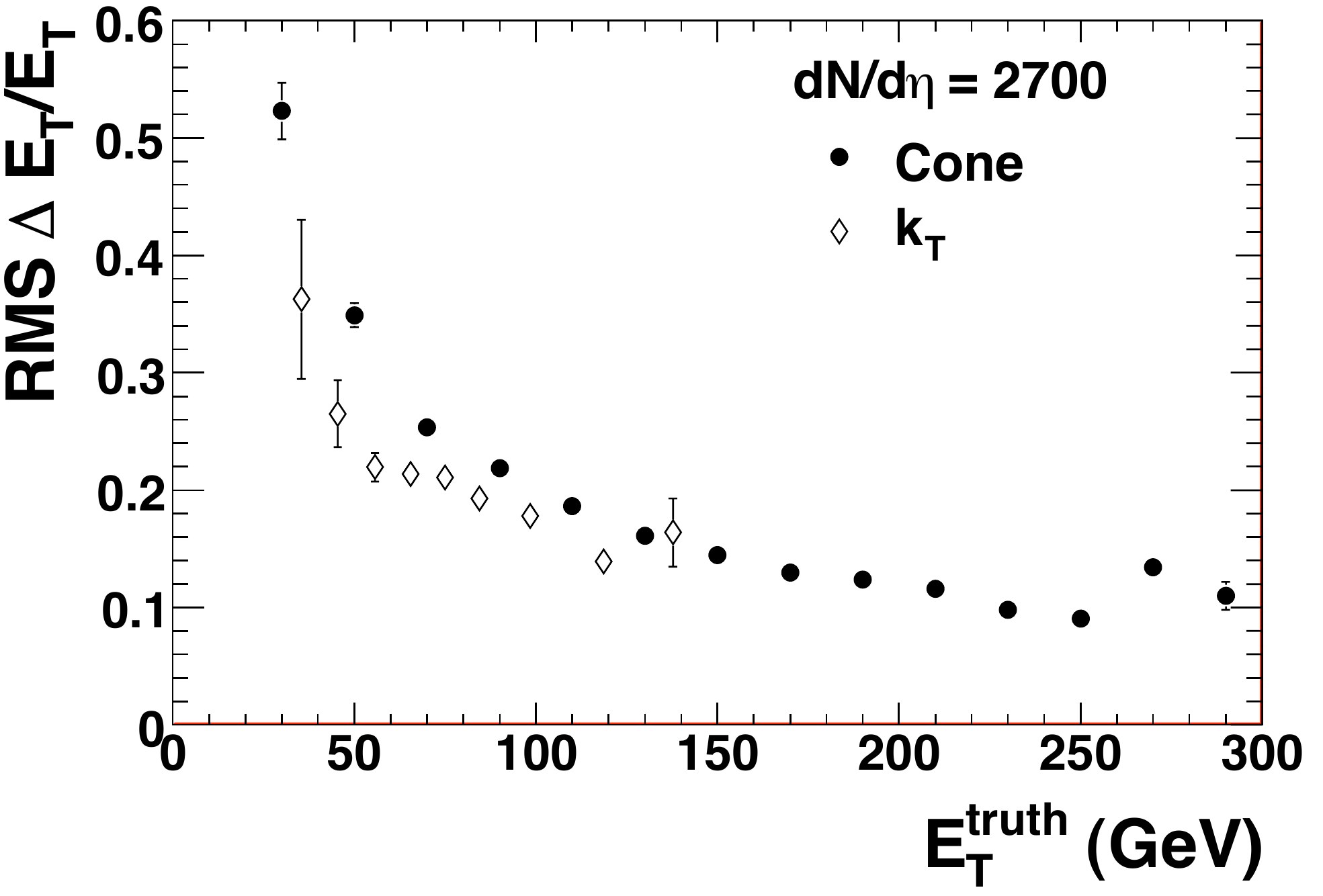}
\includegraphics[width=0.45\linewidth]{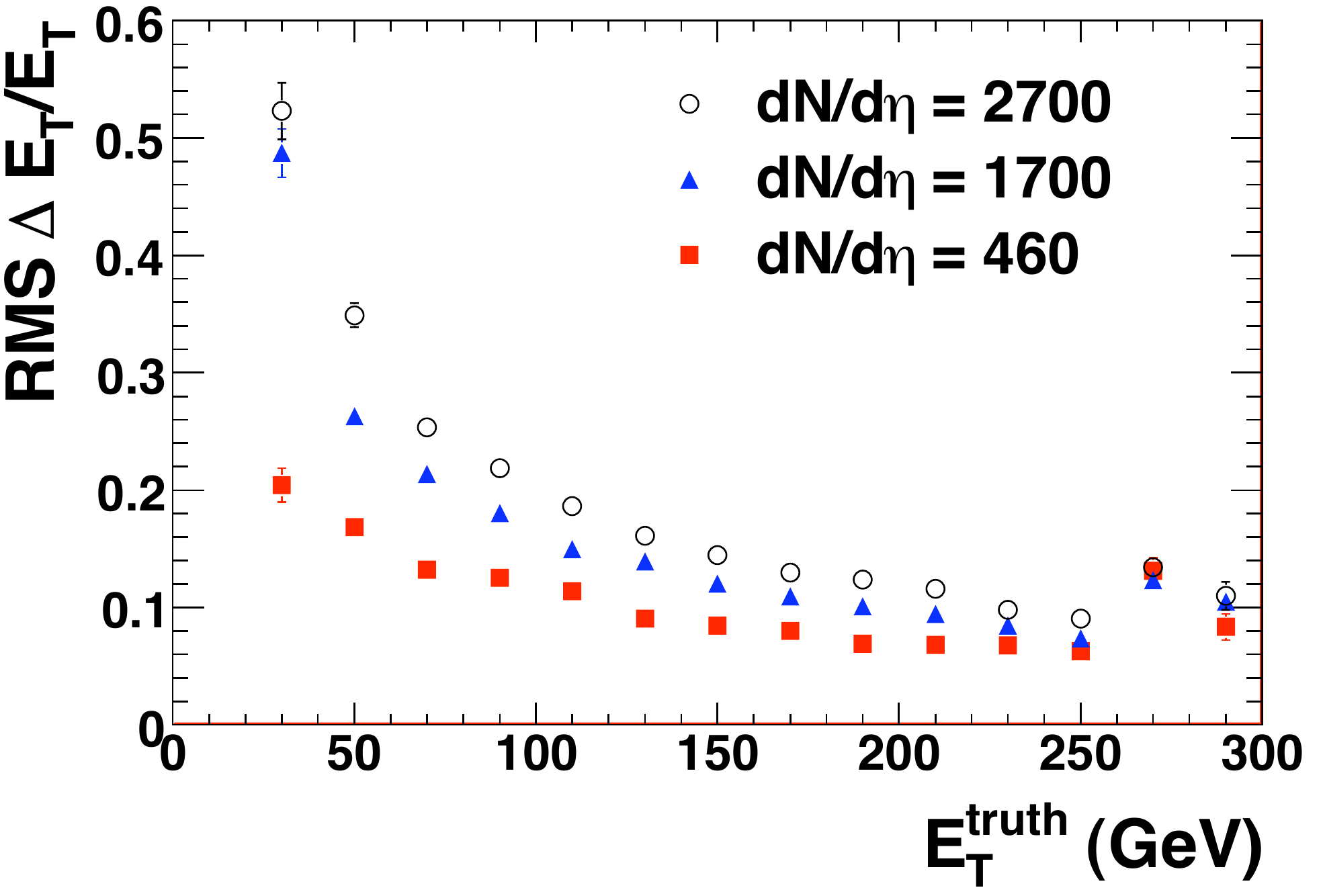}
\includegraphics[width=0.45\linewidth]{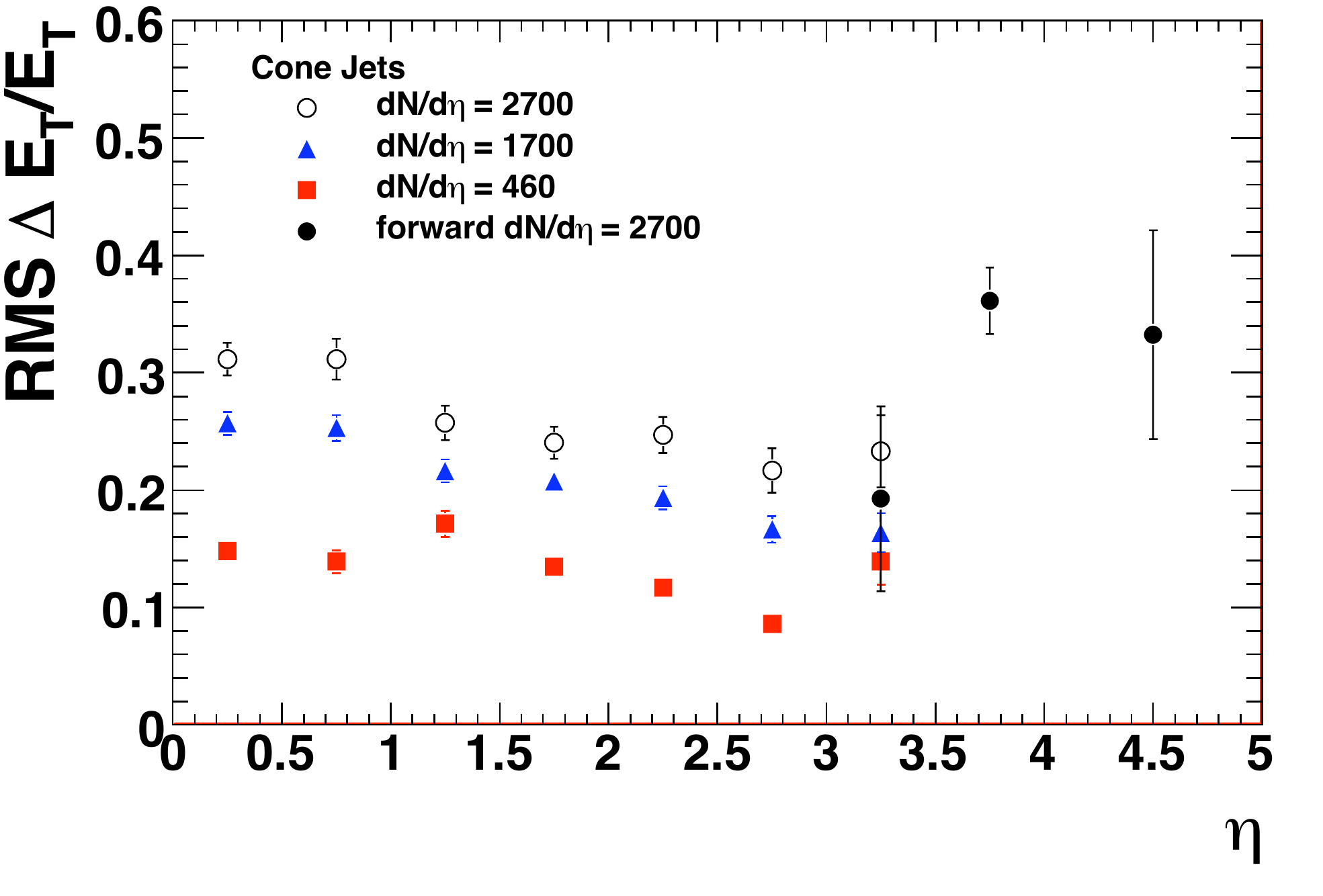}
\end{center}
\caption{
\textit{Upper:} Comparison of the jet reconstruction efficiency (left) and jet 
energy resolution (right) for cone (filled) and $k_T$ (open) algorithms from 
PYTHIA di-jet events embedded in unquenched HIJING. \textit{Lower:} Energy 
resolution of cone jets as a function of $E_T$ (left) and as a function of 
$\eta$ (right) (for $E_{T}>$ 70 GeV) for several HIJING $dN/d\eta$. The filled 
circles indicate a forward, $|\eta|>$ 3, sample of embedded jets.}\label{fig:conekt}
\end{figure}

The jet reconstruction efficiency and energy resolution for jets reconstructed 
in central, $dN/d\eta$ = 2700, events with the cone and $k_T$ algorithm are shown in 
upper panels of Fig.~\ref{fig:conekt}. Efficiency and resolution differences are 
observed for the lower $E_T$ jets.
The jet reconstruction has been studied as a function of HIJING inclusive 
charged-particle multiplicity for $|\eta|<$0.5.
The lower panels of Fig.~\ref{fig:conekt} show the energy resolution of reconstructed cone jets
as a function of $E_T$ and $\eta$. The energy resolution improves with decreasing 
multiplicity and with $\eta$
up to the region of the forward calorimeter 
(at $|\eta|>$3.2) where the resolution becomes similar to that measured at midrapidity.

\begin{figure}[tb]
\begin{center}
\includegraphics[width=0.45\linewidth]{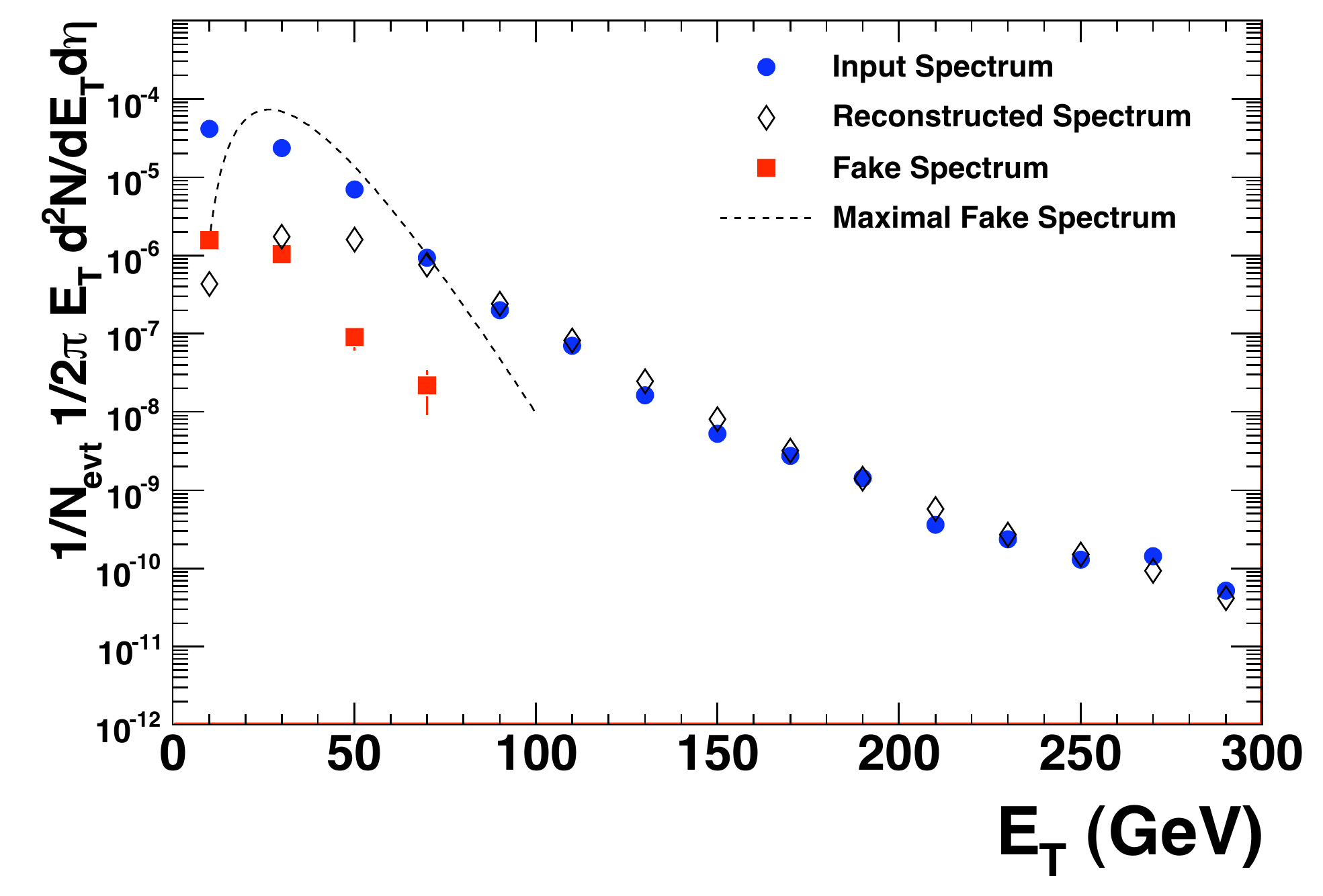}
\includegraphics[width=0.45\linewidth]{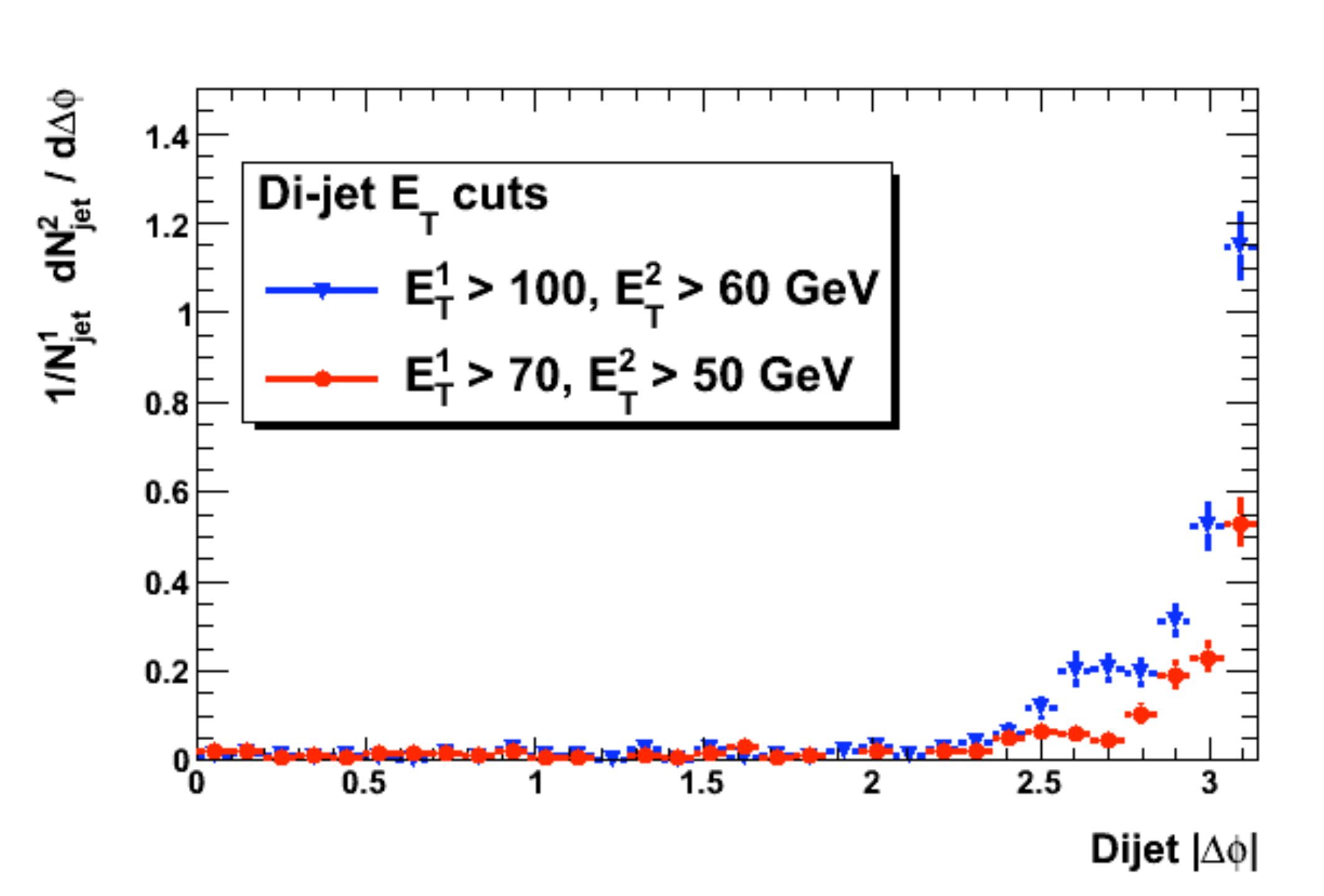}
\end{center}
\caption{\textit{Left:} Inclusive jet spectra: input PYTHIA (filled circles), 
raw reconstructed (open diamonds), raw fake jets (dashed line), and fake jets after rejection 
(squares) for cone jets reconstructed 
from PYTHIA di-jet events embedded in HIJING $dN/d\eta$ = 2700. The raw spectrum
is uncorrected for efficiency and energy resolution. \textit{Right:} The $\Delta\phi$ 
distribution between reconstructed cone jets in two different jet energy ranges.
}\label{fig:jetdijet}
\end{figure}

The reconstructed and fake cone jet spectra for $dN/d\eta$ = 2700 is shown in 
the left panel of Fig.~\ref{fig:jetdijet} and compared to the input PYTHIA jet spectrum.
The raw reconstructed spectrum is 
uncorrected for efficiency and energy resolution. Still, the raw spectrum matches 
the input distribution well for $E_T>$ 80 GeV. The maximal fake rate was 
evaluated from unquenched HIJING events. Rejection of fake jets 
was performed by making a cut on the shape of the energy distribution within the jet.
A fake fraction of less than a few percent is achieved with
minimal loss of efficiency for jets above 50 GeV.

ATLAS will also perform di-jet studies.
The $\Delta\phi$ distribution between reconstructed cone jets is shown in
the right panel of Fig.~\ref{fig:jetdijet} for two different $E_T$ cuts for the jet pairs. The
distribution is plotted as a conditional probability of observing the second, lower-$E_T$ jet given the 
trigger, higher-$E_T$ jet. This distribution has not been corrected for efficiency or
energy and position resolution. Still, integration of the higher jet $E_T$ data yields a 
60\% probability of observing a jet above 60 GeV given a jet above 100 GeV in the event. Such
a high probability before corrections reflects the high efficiency and large 
acceptance for jets in the ATLAS calorimeter.

\begin{figure}[tb]
\begin{center}
\includegraphics[width=0.45\linewidth]{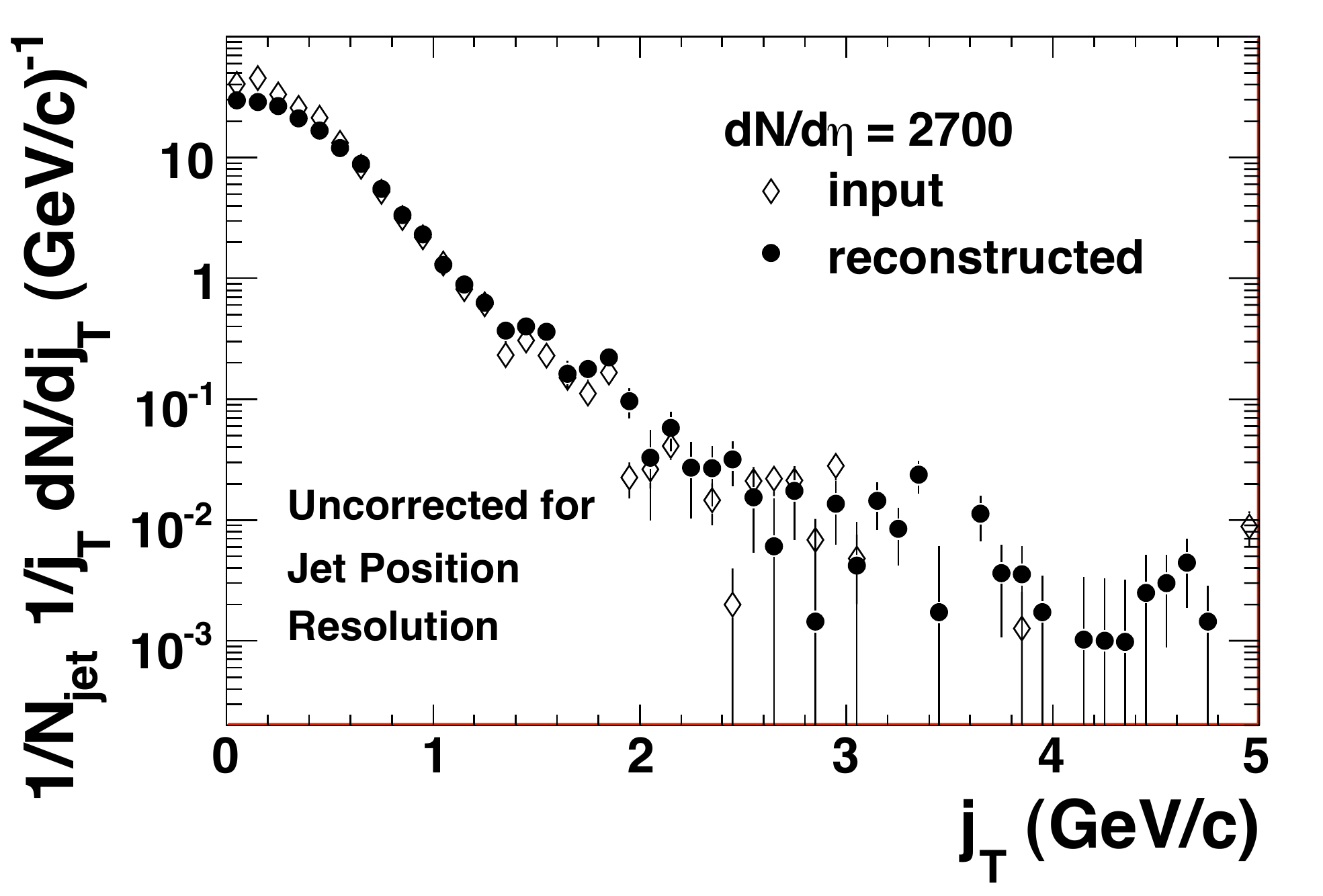}
\includegraphics[width=0.45\linewidth]{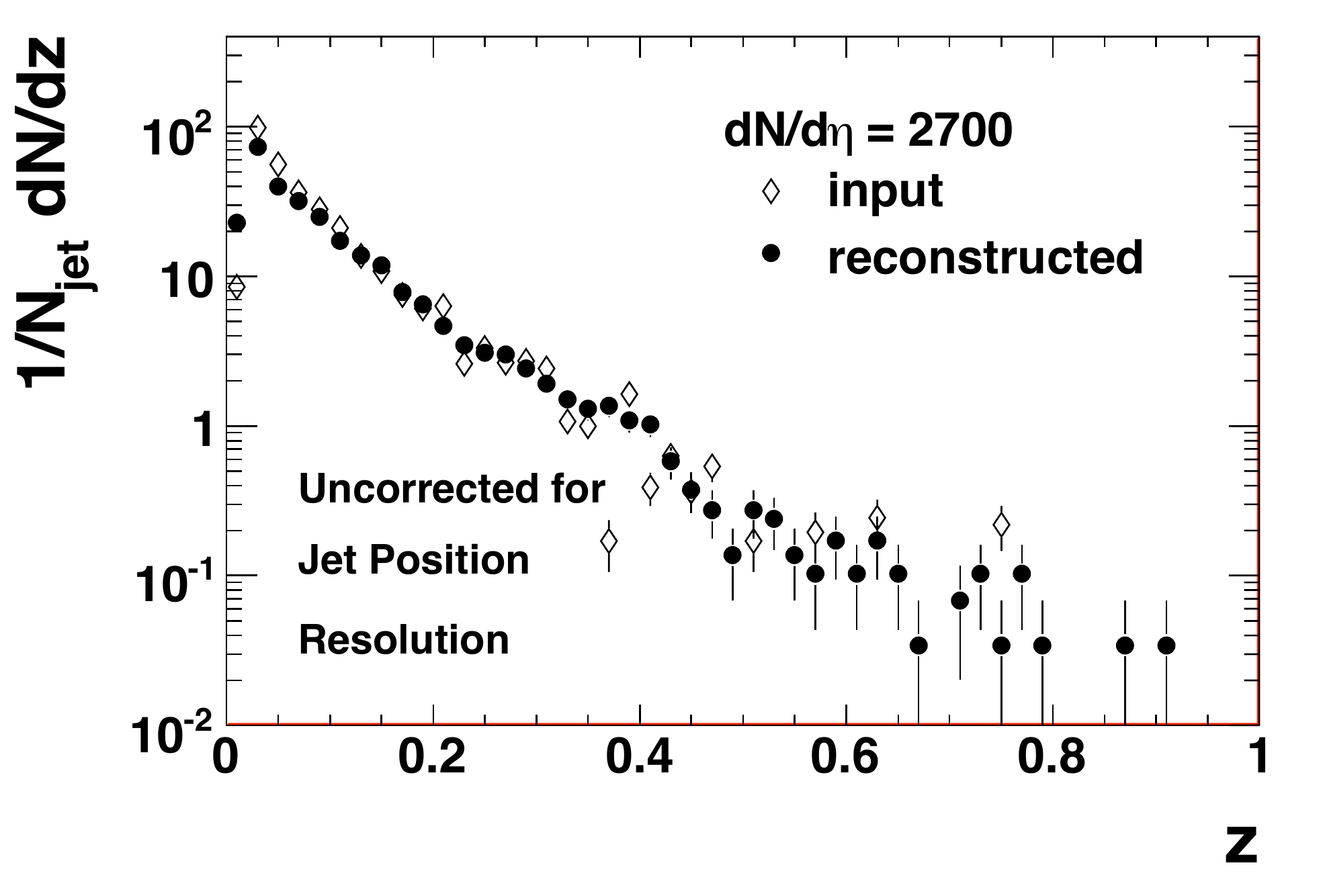}
\end{center}
\caption{The $j_T$ distribution (left) and the fragmentation function (right) from
raw reconstructed  (closed) and input PYTHIA distribution (open) for $E_T >$ 70 GeV jets. 
The raw distributions have not been corrected for jet position resolution.}\label{fig:jtdz}
\end{figure}

New and varied measurements more sensitive to energy loss will be available for study 
with fully reconstructed jets. Two examples are measurements of the fragmentation 
function, $D(z) = 1/N_{jet}$ $dN/dz$, and the $j_T$ distribution. For jets, $z$ is the
longitudinal momentum fraction of a fragment with respect to the jet and $j_T$ is 
the transverse momentum of a fragment with respect to the jet. Both distributions 
are predicted to be sensitive to details of the energy loss\cite{Salgado:2003rv,Armesto:2007dt}. 
Jet fragments are 
measured from charged tracks in the inner detector\cite{PeterQM08}, extrapolated 
to the calorimeter, and matched to the reconstructed jet. 
These distributions of charged fragments from PYTHIA jets embedded in unquenched HIJING 
events are shown in Fig.~\ref{fig:jtdz}. The open points are the PYTHIA distributions 
and the closed points are the reconstructed distributions. The latter are corrected for the 
tracking efficiency and any difference between the truth and reconstructed 
distributions are due to jet position resolution. 

\subsection{Direct Photons in ATLAS}\label{sec:photons}
\begin{figure}[tb]
\begin{center}
\includegraphics[width=0.8\linewidth]{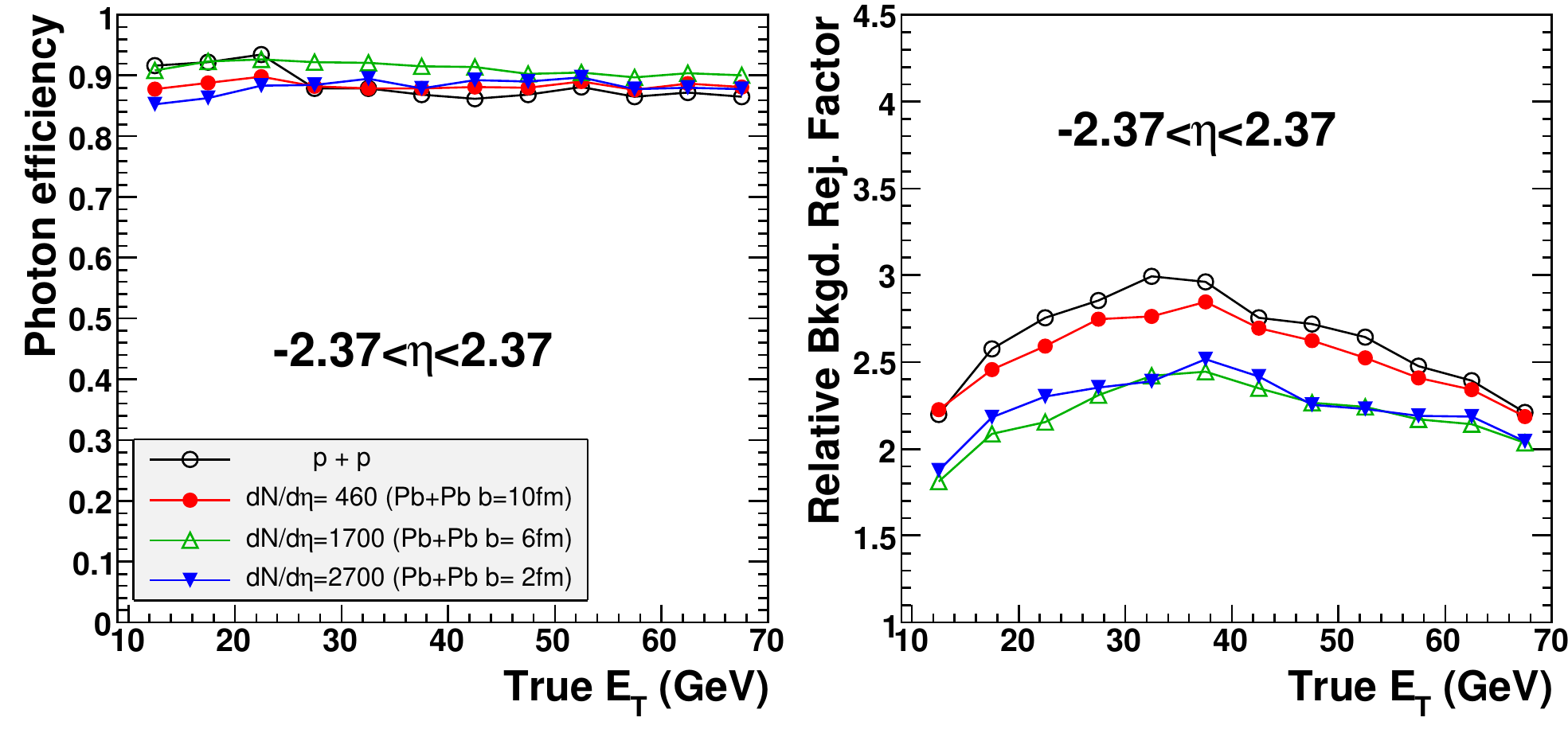}
\end{center}
\caption{Photon identification efficiency (left) and relative rejection (right) 
for photon-identification cuts from the strip layer of the calorimeter. The cuts are tuned 
to give 90\% efficiency.}\label{fig:looseid}
\end{figure}

The measurement of direct photons is another important tool for understanding
the mechanism of energy loss. Direct photons can be used as a means to pin down 
the $Q^2$ of the initial hard scattering process to study the medium modification 
of the single, recoil jet. Unfortunately, direct photons are 
amongst a large background of photons from hadronic decays. An NLO pQCD 
calculation\cite{INCNLL1} predicts a direct photon signal to decay photon background of 
2-10\% from 30-100 GeV photons at $\sqrt{s}$ = 5.52 TeV. To measure direct photons,
reconstructed electromagnetic clusters are subjected to shower shape cuts and to 
isolation criteria to suppress the substantial background from hadronic decays.

The ATLAS calorimeter was uniquely designed specifically to perform such isolation
and shape analysis for photons by rejecting di-jets for the purpose of Higgs seaches,
\textit{i.e.} $H\rightarrow\gamma\gamma$. 
The $\Delta\eta$ segmentation of the front longitudinal electromagnetic (strip) layer 
is typically about 0.003 units. With this segmentation photons from $\pi^{0}$ decays 
can be separated over a wide range of $E_T$.
Also, because of the fine segmentation,
the occupancy is quite small. In unquenched HIJING events with 
$dN/d\eta$ = 2700, the background contributes only a few hundred MeV to a cell whereas a 
few GeV photon will contribute typically half of its energy to the cell in the 
strip layer. Therefore, the efficiency of making identification cuts based on the 
information from the strip layer is only weakly dependent on centrality (see 
Fig.~\ref{fig:looseid}).

A series of shape variables can be constructed based on the energy distribution in 
the strip layer. These cuts have been tuned separately for each of the different 
HIJING multiplicity samples that have been simulated. The resulting efficiency and 
rejection of these tuned cuts for several multiplicities is shown in 
Fig.~\ref{fig:looseid}. This particular set of cuts is ``loose'' so as to keep 
the efficiency high at $\sim$90\% with a resulting relative rejection of up to 3. A set 
of ``tight'' cuts has also been tuned which results in 50\% efficiency with rejection 
up to a factor of 6. Though these rejections are not enough to overcome the 
large background from hadronic decays, it should be sufficient to perform a statistical
subtraction and extract the spectrum of non-decay photons within jets, which 
come from photon fragmentation and may be enhanced by jet-medium bremsstrahlung photons\cite{Turbide:2005fk}.

\begin{figure}[tb]
\begin{center}
\includegraphics[width=0.8\linewidth]{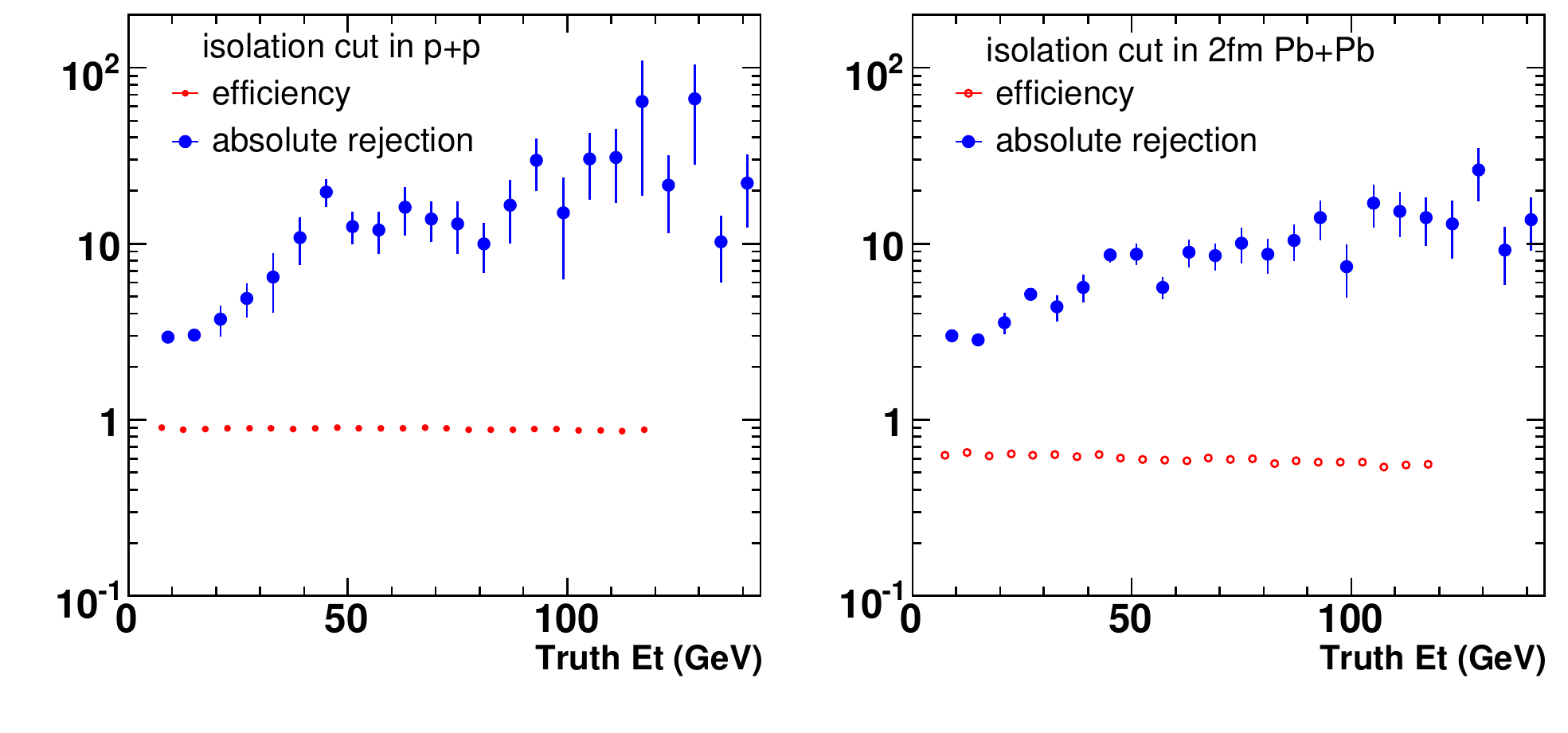}
\end{center}
\caption{The efficiency (open) and the absolute rejection (filled) of the isolation
criteria on direct photon measurements in $p+p$ (left) and $Pb+Pb$ collisions with 
$dN/d\eta$ = 2700 (right).}\label{fig:isolation}
\end{figure}

Isolation criteria based on calorimetric energy and track $p_T$ within various cone 
sizes centered on the photon have been explored to gain additional rejection. These 
isolation criteria are a strong function of centrality and have been tuned for 
different multiplicities by requiring the highest rejection with at least 50\% efficiency. 
For example, for $dN/d\eta$ = 2700, the isolation requirment within a cone of $R$=0.2 is 
$\sum E_{T} <$ 31 GeV and no reconstructed track with $p_T>$ 2.5 GeV. The rejection and 
efficiency of these cuts as a function of photon $E_T$ are shown in 
Fig.~\ref{fig:isolation}.

\begin{figure}[tb]
\begin{center}
\includegraphics[width=0.8\linewidth]{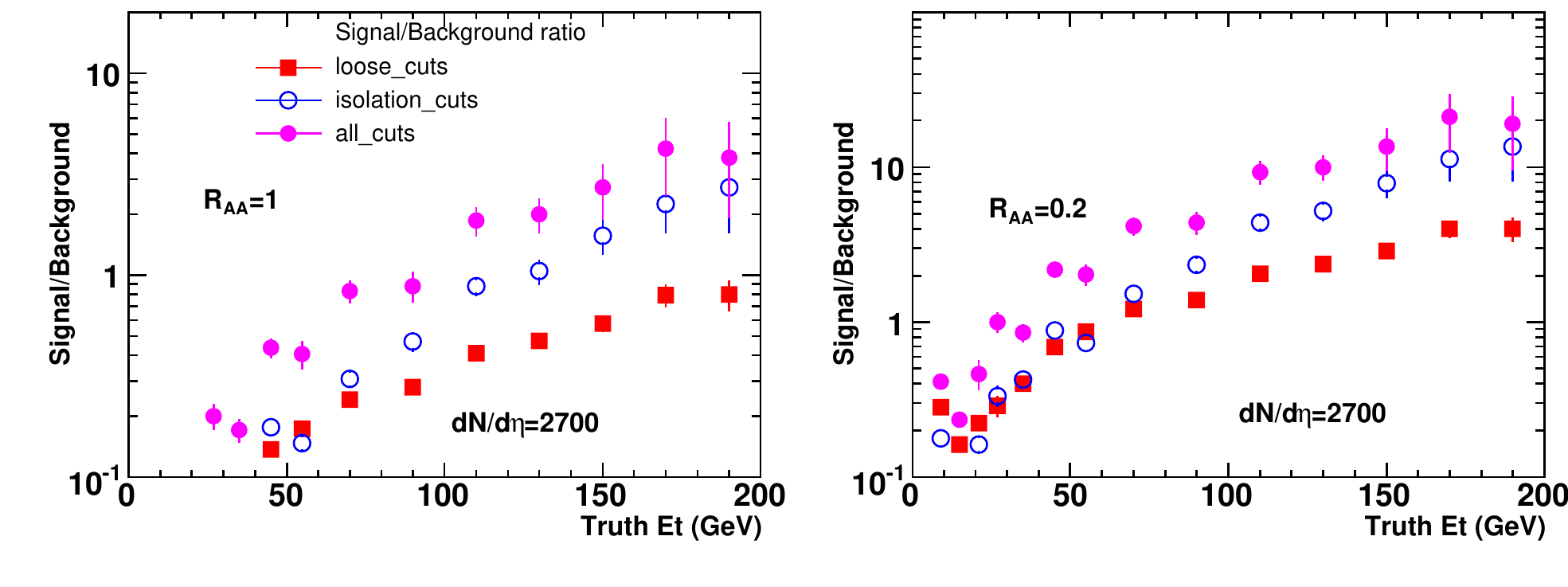}
\end{center}
\caption{
Photon signal to hadronic decay background ratio for 
unsuppressed hadrons (left) and for suppressed hadrons (right). The ratio is shown for 
the shape cuts (squares), isolation cuts (open circles), and combined (closed 
circles).}\label{fig:gammaspectrum}
\end{figure}

Combining the shape cuts with the isolation results in substantial background
rejection.
The resulting direct photon signal to hadronic decay 
background is shown in Fig.~\ref{fig:gammaspectrum}.
The ratio, with unsuppressed hadrons ($R_{AA}=1$), is $\sim$1 at 70 GeV increasing rapidly 
with $E_T$. However, if a factor of 5 suppression of hadrons from energy loss is measured 
at the LHC ($R_{AA}=0.2$), the ratio is $\sim$1 at 30 GeV. Assuming 0.5 $nb^{-1}$ of data 
per LHC year and a reconstruction efficiency of 50\%, 2x10$^5$ direct photons are expected 
with $E_T>$ 30 GeV where the signal-to-background ratio is $\gtrsim$ 1.

\section{Conclusions}
A full set of measurements are being planned to carry out the ATLAS heavy ion physics
program. Jets, photons, global observables\cite{PeterQM08}, quarkonia\cite{SashaQM08} and other 
heavy flavor physics are being explored. The uniquely designed calorimeter will make
jet and photon measurements a key strength of the ATLAS heavy ion program. The 
large acceptance and longitudinal segmentation results in high efficiency, 
more than 70\% for $E_T>$ 70 GeV, and high energy resolution, better than 25\% for $E_T>$ 70 GeV, 
at the highest expected multiplicities. Direct photons will be measured with
good efficiency, above 50\%, and large signal-to-background, S/B more than 1 for 
$E_T>$ 30 GeV, because of the uniquely designed strip layer of the electromagnetic 
calorimeter. With these measurements
ATLAS is capable of making 
key measurements which will elucidate the nature of energy loss in the QCD medium.

\section{Note}
The figures shown here were based on studies
using modified versions of ATLAS production software. Thus, they
should be considered ``ATLAS preliminary''. For completeness, 
version 12.0.6 of Athena software was used for generation, 
simulation, embedding, and reconstruction augmented only by
the jet background subtraction software discussed.


\end{document}